\documentclass[aps, pre, reprint, superscriptaddress, floatfix, longbibliography]{revtex4-2}

\usepackage[utf8]{inputenc}
\usepackage{amsmath}
\usepackage{amsfonts}
\usepackage{amssymb}
\usepackage{siunitx}
\usepackage{physics}
\usepackage{stmaryrd}
\usepackage[colorlinks=true, citecolor=blue, urlcolor=red]{hyperref}
\usepackage{xcolor}
\usepackage{orcidlink}

\newcommand{\mean}[1]{\langle {#1} \rangle}

\definecolor{mlblue}{rgb}{0, 0.4470, 0.7410}
\definecolor{K6}{rgb}{0.3010, 0.7450, 0.9330}
\definecolor{K7}{rgb}{0.6350, 0.0780, 0.1840}
\definecolor{K8}{rgb}{0, 0.4470, 0.7410}

\DeclareMathOperator{\erfc}{erfc}
\DeclareMathOperator{\e}{e}

\newcommand{\pfb}{P_{\text{FB}}}
\newcommand{\pno}{P_{\text{no}}}
\newcommand{\pRno}{P^R_{\text{no}}}
\newcommand{\trelax}{\tau_r}
\newcommand{\vfb}{V_\mathit{FB}}
\newcommand{\vdc}{V_\mathit{DC}}

\begin{document}
\title{Information engine fueled by first-passage times}

\author{Aubin Archambault\,\orcidlink{0009-0002-3373-357X}}
\affiliation{\href{https://ror.org/02feahw73}{CNRS}, \href{https://ror.org/04zmssz18}{ENS de Lyon}, \href{https://ror.org/00w5ay796}{Laboratoire de Physique}, F-69342 Lyon, France}
\author{Caroline Crauste-Thibierge\,\orcidlink{0000-0001-5502-0445}}
\affiliation{\href{https://ror.org/02feahw73}{CNRS}, \href{https://ror.org/04zmssz18}{ENS de Lyon}, \href{https://ror.org/00w5ay796}{Laboratoire de Physique}, F-69342 Lyon, France}
\author{Alberto Imparato\,\orcidlink{0000-0002-7053-4732}}
\affiliation{Department of Physics, University of Trieste, Strada Costiera 11, 34151 Trieste, Italy}
\affiliation{ Istituto Nazionale di Fisica Nucleare, Trieste Section, Via Valerio 2, 34127 Trieste, Italy}
\author{Christopher Jarzynski\,\orcidlink{0000-0002-3464-2920}}
\affiliation{Department of Chemistry and Biochemistry, Institute for Physical Science and Technology, Department of Physics, University of Maryland, College Park, Maryland 20742, USA}
\author{Sergio Ciliberto\,\orcidlink{0000-0002-4366-6094}}
\affiliation{\href{https://ror.org/02feahw73}{CNRS}, \href{https://ror.org/04zmssz18}{ENS de Lyon}, \href{https://ror.org/00w5ay796}{Laboratoire de Physique}, F-69342 Lyon, France}
\author{Ludovic Bellon\,\orcidlink{0000-0002-2499-8106}}
\email{ludovic.bellon@ens-lyon.fr}
\affiliation{\href{https://ror.org/02feahw73}{CNRS}, \href{https://ror.org/04zmssz18}{ENS de Lyon}, \href{https://ror.org/00w5ay796}{Laboratoire de Physique}, F-69342 Lyon, France}

\date{\today}

\begin{abstract}
Using a mechanical cantilever submitted to electrostatic feedback control, we investigate the thermodynamic properties of an information engine that extracts work from thermal fluctuations. The cantilever position is rapidly sampled and the feedback is triggered by the first passage of the system across a fixed threshold. The information $\Delta I$ associated with the feedback is based on the first-passage-time distribution. In this setting, we derive and experimentally verify two distinct fluctuation theorems that involve $\Delta I$ and give a tight bound on the work produced by the engine. Our results extend beyond the specific application to our experiment: we develop a general framework for obtaining fluctuation theorems and work bounds, formulated in terms of probability distributions of protocols rather than underlying measurement outcomes.
\end{abstract}

\maketitle

Information plays a fundamental role in the thermodynamics of mesoscopic systems, where thermal fluctuations cannot be ignored~\cite{Rex, Parrondo_sagawa, CiliLutz_PT, CiliLutzchap}. The erasure of information costs energy~\cite{Parrondo_sagawa, CiliLutz_PT, CiliLutzchap}, by Landauer's principle~\cite{Landauer1961}. Conversely, heat can be converted into work by an engine that uses, as fuel, information gathered through measurements, realizing in this way a Maxwell demon~\cite{, Bennett1982,Bechoefer_book_2021, Rex}. Two central issues in the study of these engines are: (i) identifying relevant measures of information, and (ii) determining bounds on the amount of work the engine can produce. The Sagawa-Ueda equality~\cite{sagawa_generalized_2010} relates the mutual information~\cite{Cover2006}, $I$, between the system variable and the measurement outcome, to the work performed, $w$, and the free energy difference, $\Delta F$:

\begin{equation}
	\mean{\\e^{- w- I}} = \e^{- \Delta F},
	\label{eq:GJE}
\end{equation}
where $\mean{\cdot}$ is an ensemble average over independent realizations of the process. Throughout this paper, $w$, $\Delta F$ and all other energetic quantities are expressed in units of $k_BT$, where $k_B$ is Boltzmann's constant and $T$ denotes temperature. Eq.~\ref{eq:GJE} combines with Jensen's inequality~\cite{Chandler87} to give an upper bound on the mean extracted work:
\begin{equation} \label{EqWbound}
 \mean{-w}\le \mean{I}-\Delta F.
\end{equation}
Several feedback protocols have been proposed in theoretical models~\cite{Horowitz_PRE_2010,Horowitz_2011, Seifert_entropy_feddback_2012, Lahiri_2012, Rosinberg_EPL_2016, ashida_general_2014} and in experiments~\cite{Toyabe2010, Pekola_mutual_information_2014, Koski_2015, Camati_2016, Chida_2017, Roichman_2018, PRL_Pak_exprment_2018, Ribezzi_2019, IEEE-2022, Bechhoefer_2022}, in order to optimize the power that such information engines deliver, and to test bounds such as Eq.~\ref{EqWbound} or stronger fluctuation theorems such as Eq.~\ref{eq:GJE}.

In Ref.~\onlinecite{sagawa_generalized_2010}, Eq.~\ref{eq:GJE} is derived by assuming that measurements are performed with errors. In the absence of measurement errors, the mutual information $\langle I \rangle$ reduces to the Shannon entropy~\cite{Toyabe2010,Ribezzi_2019} of the measurement outcome, and Eq.~\ref{eq:GJE} can be violated. A subtlety arises for continuous system variables, such as position, $x$, which are necessarily observed with a finite bin size, $\delta x$. In this case the Shannon entropy contains a term that scales as $-\log(\delta x)$~\cite{marsh2013introduction}. Thus, while Eq.~\ref{eq:GJE} can be violated for error-free measurements, Eq.~\ref{EqWbound} becomes satisfied trivially, and loosely, by taking $\delta x$ to be sufficiently small. The divergence of the Shannon entropy can be surmounted by measuring the entropy production rate~\cite{Roichman_2018}, but the bound remains loose.

To overcome the divergence of $I$, to allow for error-free measurements, and to obtain a bound tighter than Eq.~\ref{EqWbound}, Ashida et al.~\cite{ashida_general_2014} derived the following equality:
\begin{equation}
\label{eq:ashida}
 \mean{\e^{-w+\Delta F-I+I_u}} = 1.
\end{equation}
Here, the \emph{unavailable information}, $I_u$, is a function of the set of measurement outcomes during a realization of the process. $I_u$ quantifies the information that is measured but unused in the feedback protocol, assuming error-free measurements. The extracted work is bounded by $\Delta I\equiv I-I_u$, which no longer contains a binning term $-\log(\delta x)$. Eq.~\ref{eq:ashida} has been extended to measurements made with errors; in this case $I$ is again the mutual information, but the work bound is tighter than Eq.~\ref{EqWbound}~\cite{Potts_PRL_2018}. Although Eq.~\ref{eq:ashida} addresses the issue related to the binning term, it requires the analysis of stochastic trajectories with a time-reverse protocol, which is often not accessible experimentally. In Ref.~\onlinecite{Archambault-EPL} both $I$ and $I_u$ were measured, but only for a limited regime where all measurements are made in equilibrium.

In this Letter, we study an information engine fueled by the rapid sampling of an underdamped Brownian particle's position. We derive a fluctuation theorem formally equivalent to Eq.~\ref{eq:ashida}, but with $I$ and $I_u$ \emph{based on protocols} rather than measurement outcomes. We demonstrate that both $I$ and $I_u$ can be accessed experimentally, and that the resulting bound on the extracted work is saturated. We also derive a new relationship between work and information:
\begin{equation}
	\label{eq:newFT}
	\mean{\e^{-w+\Delta F}} = \mean{\e^{I-I_u}}.
\end{equation}
The right side of Eq.~\ref{eq:newFT} is the counterpart of the {\it efficacy}, $\gamma$, introduced in Ref.~\cite{sagawa_generalized_2010} for feedback with measurement errors, that can be measured and expressed simply thanks to our definition of information. Our derivations, presented below for our experiment, are generalized to arbitrary feedback protocols in the End Matter (EM).

\begin{figure}[tb]
\centering
\includegraphics[width=\linewidth]{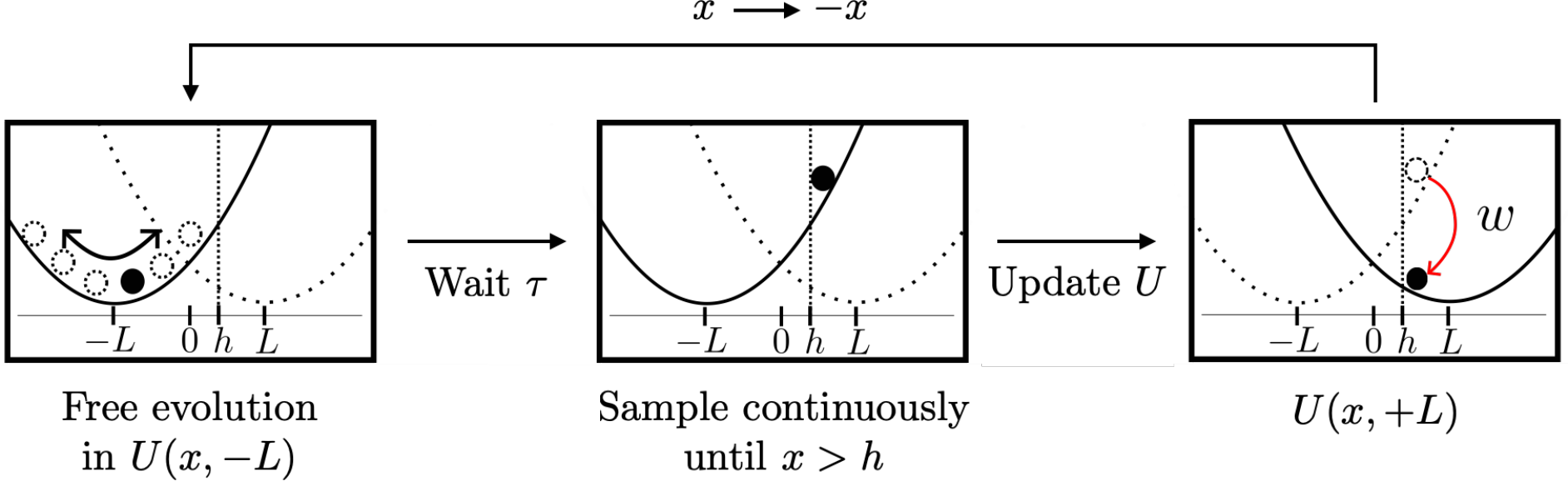}
\caption{\label{Fig:prot_cont} First-passage protocol. Initially, the demon is locked and the bead equilibrates in the potential $U_A(x)=U(x, -L)=\frac{1}{2}(x+L)^2$ for a time $\tau\gg \trelax$. The demon is then activated and rapidly samples the bead position $x$. As soon as $x>h$ (which might occur at the first sampling), the potential is switched to $U_B(x)=U(x, +L)=\frac{1}{2}(x-L)^2$, the demon is locked and the protocol is repeated (with a symmetry $x\leftrightarrow-x$ to recover the same initial state).}
\end{figure}

Our information engine -- inspired by the one introduced by Sagawa and Ueda (SU protocol)~\cite{sagawa_generalized_2010} and tested by Toyabe {\it et al}~\cite{Toyabe2010} -- is illustrated in Fig.~\ref{Fig:prot_cont}. A 1D Brownian particle begins in equilibrium, in a harmonic potential $U(x, -L)=\frac{1}{2} (x+L)^2$. An external agent, or ``demon'', monitors the particle's position, $x$, every short time step $\delta t$, and compares it to a threshold $h>0$. As soon as $x>h$, the demon shifts the potential's center from $-L$ to $+L$, extracting an amount of work $-w=U(x, -L)-U(x, L)=2 L x$. The demon is then ``locked'': the particle equilibrates, evolving undisturbed for a time $\tau$ much longer than its relaxation time $\trelax$. The demon is then unlocked, and a new engine cycle begins. For experimental convenience, the new cycle is a symmetric image of the previous cycle, using $-x$ instead of $x$; thus, from cycle to cycle the location of potential's minimum toggles between $-L$ and $+L$~\footnote{As the system begins at equilibrium in the harmonic potential, this toggling does not affect the statistics of work or information.}. During each engine cycle, the particle remains in equilibrium until the demon is triggered as the threshold $h$ is crossed. However, since measurements of $x$ are made in rapid succession (small $\delta t$), they are strongly correlated, making it challenging to compute Shannon information $I$ and unavailable information $I_u$ defined in Ref.~\cite{ashida_general_2014}.

We have realized this protocol experimentally using an underdamped oscillator, as described in Ref.~\onlinecite{Archambault-EPL,Dago-2021, Dago-2022-JStat,Dago-chapter} and in the Supplemental Material (SM) section \ref{SMA}. In a nutshell, the oscillator is a cantilever subject to thermal noise, whose position $x$ is measured precisely with an interferometer. The standard deviation $\sigma$ of $x$ in thermal equilibrium sets the unit of length. The position of the well's center, $\pm L$, is set by an external electrostatic force, driven by a feedback loop following the protocol described above. The feedback response time is several orders of magnitude shorter than any characteristic time of the oscillator, and can be neglected. An analogous protocol, triggered by the unfolding of DNA, was used to reduce dissipation in single-molecule stretching experiments~\cite{rico-pasto2021}.
Information engines triggered by first-passage times are also realized experimentally in Refs.~\cite{Roichman_2018,Ribezzi_2019,Bechhoefer_2022}.

We performed experiments for a range of values of $h$ and $L$, always using $\tau = 5 \trelax$ to generate a fresh equilibrium state after each cycle of work extraction. At each value of $h$ and $L$, we record for a few minutes at a sampling time $\delta t=\SI{0.5}{\mu s}$: the position, $x$, the state of the demon (locked or active) and the center of the trap, $\pm L$. We extract for each trigger of the demon the time $t_k=k\delta t$ and the position $x(t_k)$ at which it occurred. The stochastic work measured is simply $w=-2Lx(t_k)$, and we record the values of $w$ and $k$ to evaluate their statistics.

\begin{figure}[b]
\centering
\includegraphics{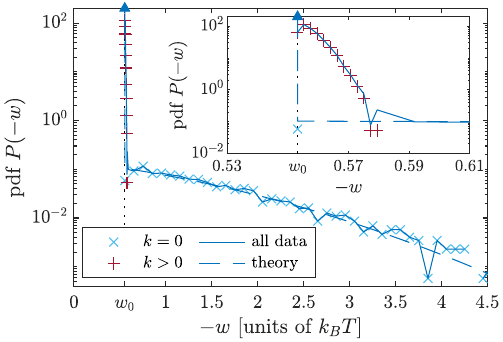}
\caption{\label{Fig:pdf_FB1} Probability distribution function (pdf) of extracted work, $P(-w)$, for $L=0.91$ and $h=0.30$. For each event, the number of
	 readings $k$ performed before the switching is measured. The full pdf consists of a peak near $w_0=2Lh$ from events $k>0$ (\textcolor{K7}{$+$}) and a tail from events $k=0$ (\textcolor{K6}{$\times$}), as expected from theory. The finite sampling time $\delta t$ produces the spread around $w_0$ (rather than a Dirac distribution, \textcolor{K8}{$\blacktriangle$}) shown in the inset, as the trigger does not
	 occur exactly at $x(t_k)=h$.
}
\end{figure}

As an example, Fig.~\ref{Fig:pdf_FB1} shows the probability distribution function (pdf) of the extracted work $-w$ during an experiment performed at $L=0.91$ and $h=0.30$. The pdf can be decomposed into two contributions by conditioning on the value of $k$. The prominent peak at $w_0=2Lh$ corresponds to events for which $k>0$, that is, $x(t_{k-1})\le h < x(t_{k})$. In this case $-w=2Lx(t_k)\simeq 2Lh\equiv w_0$, since the system barely moves during the sampling time $\delta t=t_{k}-t_{k-1}$. In contrast, the tail at $-w>w_0$ represents all trajectories for which $k=0$. In this case the system begins at $x(t_0)>h$, triggering the demon at the first reading. Since $w=-2Lx(t_0)$ and $x(t_0)$ is sampled from equilibrium, this tail is easily computed (Eq.~\ref{eq:pdfw}) and perfectly matches the measurement.

\begin{figure}
 \centering
 \includegraphics{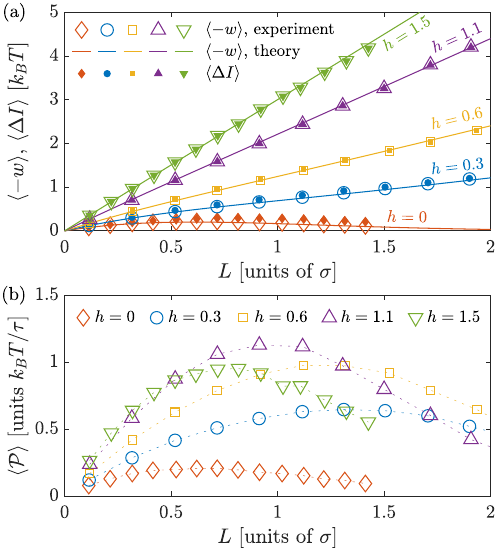}
 \caption{(a) Extracted mean work $\mean{-w}$ as a function of $L$ and $h$: experimental results (open markers, statistical uncertainty smaller than the symbol size), theoretical prediction (solid lines), and mean information upper bound $\mean{\Delta I}$ (filled markers, computed with Eqs.~\ref{eq:Delta_I1} and \ref{eq:Delta_Isup1}). The bound $\mean{-w}\le\mean{\Delta I}$ is verified and nearly saturated in all the measurements: the efficiency $\mean{-w}/\mean{\Delta I}$ tends to 1 when $L$ and $h$ are large. In that limit, the demon is rarely triggered at the first measurement, thus $-w=\Delta I=w_0$ for most realizations. (b) Extracted mean power as a function of $L$ and $h$.
 The maximum power is reached for $L \sim h \sim 1$.}
 \label{Fig:bound_W}
\end{figure}

For given parameter values $L$ and $h$, we compute the mean work per trigger $\mean{-w}$ and the mean power $\mean{\mathcal{P}}$ that the demon extracts during the operation. The results are plotted in Fig.~\ref{Fig:bound_W}, where we report experimental results and the theoretical value for $\mean{-w}$ (Eq.~\ref{eq:meanW}). We observe that $\mean{-w}$ generally increases with $h$ and $L$ in the explored range (except for $h=0$): the extracted work is always greater than $w_0=2Lh$. The power $\mean{\mathcal{P}} = \mean{-w}/(\tau+\mean{t_k})$ presents a maximum value when exploring the parameter space $(L,h)$: high power requires large and frequent work extraction, the latter criterion failing for large values of $L$ and $h$, as the first passage time of the engine increases. The transformation performed is simply a translation of a harmonic well, thus $\Delta F=0$ and Eq.~\ref{eq:ashida} imposes $-\mean{w}\leq \mean{\Delta I}$.

Our engine uses measurement outcomes ${\vec m}\equiv (m_0,m_1, \cdots \, m_M)$, where each $m_n$ is a binary variable that records whether the threshold $h$ has been crossed at the time of the $n$'th measurement.
As the measurements are error-free, our engine violates Eq.~\ref{eq:GJE} but satisfies Eq.~\ref{eq:ashida}.
In Ref.~\cite{ashida_general_2014}, $I({\vec m})$ and $I_u(\vec m)$ are defined in terms of probabilities of obtaining outcomes $\vec m$ during forward and reverse processes.
When the particle's position is sampled frequently, determining these probabilities becomes unfeasible, both because the number of possible outcomes $\vec m$ grows exponentially with the number of measurements, $M$, and due to strong correlations between the $m_i$'s.
We therefore formulate an alternative approach, in which $I$ and $I_u$ are given in terms of protocols rather than measurement outcomes.

We consider a stochastic system evolving in a potential $U_\lambda(x)$, where $x$ is the microscopic coordinate and $\lambda$ is a parameter that is manipulated by the demon (i.e. the feedback control). In the following $\lambda$ takes two values, $A$ and $B$. While our analysis involves a 1D system, more generally $x$ can be multidimensional (see EM). Starting with the system at equilibrium in the potential $U_A(x)$, the position of the bead $x_0, x_1\ldots x_M$ is measured at discrete times $t_n = n\times \delta t$, with $\delta t$ arbitrarily small. The measured position is compared at each time step with a threshold $h$, producing the binary sequence $m_0, m_1\ldots m_M$. Immediately after the first instant $t_k$ satisfying $x_k>h$, the parameter $\lambda$ is switched from $A$ to $B$ and then kept at $B$ independently of the following measurements. The final measurement is made at time $t_M$. 

This process produces a protocol $\Lambda= (\lambda_0, \lambda_1, \ldots , \lambda_M)$ and a trajectory $X=(x_0, x_1, \ldots , x_M)$. Here, $\lambda_0=A$ and $\lambda_{n>0}$ is the value of $\lambda$ during the interval $(t_{n-1},t_n]$. The protocol has the form:
\begin{equation} \label{eq:defLambda_k}
\Lambda = (\underbrace{A, A, \ldots, A}_{k+1}, \underbrace{B, \ldots, B}_{M-k}) = \Lambda_k, 
\end{equation}
with a sequence of $k+1$ initial $A$'s followed by $(M-k)$ $B$'s, determined by the first time $t_k$ where $x_k>h$.
We group the trajectories into sets $\Omega_k$ containing all the trajectories that first cross the threshold $h$ at time $t_k$:
\begin{equation}
\Omega_k = \lbrace X \vert \min\lbrace n \vert x_n>h\rbrace = k \rbrace.
\end{equation}
The integer variable $k$ labels the protocol, $\Lambda_k$.
For each protocol, we now define an information, $I(k)$, and an unavailable information, $I_u(k)$, analogous to the quantities $I(\vec m)$ and $I_u(\vec m)$ of Ref.~\onlinecite{ashida_general_2014}.

Let $\pfb(X, \Lambda_k)$ denote the joint probability to obtain trajectory $X$ and protocol $\Lambda_k$ when performing the process described above. By construction,
\begin{equation}\label{eq:P0_omegak}
\pfb(X, \Lambda_k) = 0, \ \text{if} \ X \notin \Omega_k.
\end{equation}
Let $P_k$ be the probability of obtaining the protocol $\Lambda_k$:
\begin{equation}
P_k = \sum_X \pfb(X, \Lambda_k) = \sum_{X\in\Omega_k} \pfb(X, \Lambda_k). 
\end{equation}
Then $\pfb(X\vert\Lambda_k) = \pfb(X, \Lambda_k) / P_k$ is the probability to obtain the trajectory $X$, conditioned on obtaining the protocol $\Lambda_k$.

Now consider a different scenario: choose a protocol $\Lambda_k$ and apply it without feedback. Let $\pno(X\vert\Lambda_k)$ denote the probability of obtaining a trajectory $X$ under this no-feedback protocol $\Lambda_k$. It follows that
\begin{equation}\label{eq:PfbPno}
\pfb(X, \Lambda_k) = \pno(X\vert\Lambda_k) \ \text{if} \ \ X\in\Omega_k.
\end{equation}
Explicitly, we have
\begin{eqnarray}
	\pno(X\vert\Lambda_k)& = &\pi_A(x_0) \cdot p_A(x_1\vert x_0) \cdots \nonumber \\ 
	& & p_A(x_k \vert x_{k-1}) \cdot p_B(x_{k+1} \vert x_k) \nonumber \\
	& & \cdots p_B(x_M\vert x_{M-1}), 
 \label{pno:def}
\end{eqnarray}
where $\pi_\lambda$ is the equilibrium distribution in potential $U_\lambda$ and $p_\lambda(x'\vert x)$ is the transition probability from $x$ to $x'$, in time $\delta t$, with control parameter $\lambda$. Eq.~\ref{pno:def} holds both when $X \in \Omega_k$ and when $X \not\in \Omega_k$. If $X \in \Omega_k$, then $\pfb(X, \Lambda_k)$ has the same expression as the right side of Eq.~\ref{pno:def}, which establishes Eq.~\ref{eq:PfbPno}.

Let $Q_k$ denote the probability to obtain a trajectory in $\Omega_k$ when imposing a protocol $\Lambda_k$ without feedback:
\begin{equation} \label{eq:DefQk}
Q_k = \sum_{X \in \Omega_k} \pno (X\vert \Lambda_k).
\end{equation}
Summing Eq.~\ref{eq:PfbPno} over all $X\in\Omega_k$, we have
\begin{equation}\label{eq:PkQk}
P_k = Q_k.
\end{equation}

Given a protocol $\Lambda_k = (\lambda_0, \ldots, \lambda_M)$, let $\Lambda_k^\dagger = (\lambda_M, \ldots, \lambda_0)$ denote the reverse protocol and $\pRno(X^\dagger\vert \Lambda_k^\dagger)$ the probability of observing a trajectory $X^\dagger$, which is the time--reversed image of $X$, when applying protocol $\Lambda_k^\dagger$ without feedback. We define $Q_k^R$ similarly to $Q_k$:
\begin{equation} \label{eq:DefQkR}
 Q_k^R = \sum_{X\in\Omega_k} \pRno(X^\dagger\vert \Lambda_k^\dagger).
\end{equation}
$Q_k$ is the probability that a trajectory generated using protocol $\Lambda_k$, without FB, first crosses the threshold after $k$ steps. $Q_k^R$ is the probability that the trajectory generated using $\Lambda_k^R$, without FB, {\it last} crosses the threshold in the reverse direction after $M-k$ steps. In Eqs.~\ref{eq:DefQk} and \ref{eq:DefQkR}, each term in the sum is defined with respect to the same protocol, $\Lambda_k$ or $\Lambda_k^\dagger$. By Eq.~\ref{eq:PkQk}, $Q_k$ sums to unity, but the same is not true of $Q_k^R$. From Crooks's theorem~\cite{Crooks_PRE} we have: 
\begin{equation}\label{eq:Crooks_FB1}
\pno(X\vert\Lambda_k)=\pRno(X^\dagger\vert\Lambda_k^\dagger)\e^{w(X, \Lambda_k)-\Delta F},
\end{equation}
where $w(X, \Lambda)$ is the work performed on the system along a trajectory $X$ under protocol $\Lambda$, and $\Delta F$ is the free energy difference between potentials $U_A$ and $U_B$. 

We now define the information and its unused part:
\begin{equation}
 I(k)= -\ln Q_k, \quad I_{u}(k)=-\ln Q_k^R.
\end{equation}
These definitions are formally similar to Eqs.~3 and 5 of Ref.~\onlinecite{ashida_general_2014}, but there is a crucial difference.
In our first-passage protocol, the particle's position is measured frequently, rendering the construction of $I(\vec m)$ and $I_u(\vec m)$ impractical.
By contrast, $I(k)$ and $I_u(k)$ can be determined readily, as we show later.
Taking
\begin{equation} \label{EqDefDI}
 \Delta I = I - I_u = -\ln \left(Q_k/Q_k^R\right),
\end{equation}
we derive Eq.~\ref{eq:ashida} (see Eq.~\ref{eqproof_ahida}):
\begin{equation}
\mean{ \e^{-w+\Delta F-\Delta I}}_\mathrm{FB}= \sum_{\Lambda_k} Q_k = \sum_{\Lambda_k} P_k = 1.
\label{eq:Expavg}
\end{equation}
This result is the counterpart of Eq.~12 of Ref.~\onlinecite{ashida_general_2014}, but using $\Delta I(k)$, based on protocols, rather than $\Delta I(\vec m)$, based on measurement outcomes.
For our engine, there are $2^{M+1}$ possible sets of measurement outcomes, $\vec m$, but only $M+1$ protocols, $k$. Below, we compute $\Delta I(k)$ analytically (Eqs.~\ref{eq:Delta_I1}, \ref{eq:Delta_Isup1}).

We additionally have, for each protocol $\Lambda_k$ (see Eq.~\ref{eq:EMCondAvg_k} for the proof),
\begin{equation}
\mean{\e^{-w+\Delta F}}_{\mathrm{FB}, k} = \frac{Q^R_k}{Q_k} = \e^{\Delta I(k)}. \label{eq:CondAvg_k}
\end{equation}
Averaging Eq.~\ref{eq:CondAvg_k} over all values of $k$, we prove Eq.~\ref{eq:newFT}:
\begin{align} 
\mean{\e^{-w+\Delta F}}_\mathrm{FB} &= \sum_k P_k \mean{\e^{-w+\Delta F}}_{\mathrm{FB},k} \nonumber \\
& = \sum_k P_k \e^{ \Delta I(k) } = \mean{\e^{\Delta I}}_\mathrm{FB}. \label{eq:corrolary}
\end{align}
Eqs.~\ref{eq:CondAvg_k} and \ref{eq:corrolary}  relate work fluctuations during a process {\it with} feedback, to the ratio of probabilities $Q_k^R/Q_k$ of outcome $k$ during processes {\it without} feedback.

Combining Eqs.~\ref{eq:PkQk} and \ref{EqDefDI} with the second line of Eq.~\ref{eq:corrolary}, we obtain the equivalent result,
\begin{equation}
\label{eq:sumQk}
 \mean{\e^{-w+\Delta F}}_\mathrm{FB} = \sum_k Q_k^R ,
\end{equation}
consistent with our earlier statement that $Q_k^R$ (unlike $Q_k$) does not generally sum to unity. Eq.~6 of Ref.~\onlinecite{sagawa_generalized_2010} is the analogue of Eq.~\ref{eq:sumQk} for measurements with errors. Our proof can be extended to arbitrary error-free feedback protocols as detailed in the EM.

We now return to the harmonic potential with the feedback control depicted in Fig.~\ref{Fig:prot_cont}, with $\Delta F = 0$. As intuited from the experimental data, we distinguish the cases $k=0$ and $k>0$. For all trajectories where $k=0$, the first measured position is already beyond the threshold: $x_0>h$. Since this measurement occurs in equilibrium, we can compute $\Delta I$ analytically (Eqs.~\ref{eq:EMQ0}-\ref{eq:EMDelta_I1}):
\begin{equation}
\Delta I(k=0) = -\ln\left( \frac{\erfc\left(\frac{h+L}{\sqrt{2}}\right)}{\erfc\left(\frac{h-L}{\sqrt{2}}\right)} \right). \label{eq:Delta_I1}
\end{equation}
When $k>0$, the potential is switched just as the bead crosses the threshold, $x=h$, and the extracted work is therefore $-w = w_0 \equiv 2L h$. Hence, $\mean{\e^{-w}}_k= \e^{2Lh}$ for all $k>0$. Eq.~\ref{eq:CondAvg_k} then implies
\begin{equation}
\Delta I(k>0) = 2Lh. \label{eq:Delta_Isup1}
\end{equation}
In SM section \ref{SMB}, we establish this result independently of Eq.~\ref{eq:CondAvg_k}.

\begin{figure}[!t]
\centering
\includegraphics{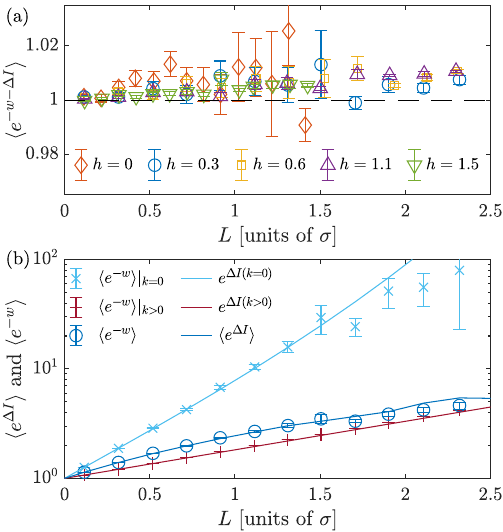}
\caption{\label{Fig:JarzFB1} (a) $\mean{\e^{-w-\Delta I}}$ as a function of $L$ for different values of $h$. As predicted by Eq.~\ref{eq:ashida}, this average is close to $1$ for all values of $L$ and $h$. (b) $\mean{\e^{-w}}$ for $k=0$ (\textcolor{K6}{$\times$}), $k>0$ (\textcolor{K7}{$+$}) and over all $k$ (\textcolor{mlblue}{$\circ$}) as a function of $L$, for $h=0.3$. As predicted by Eq.~\ref{eq:newFT} (for all $k$) and Eq.~\ref{eq:CondAvg_k} (for any specific $k$) these measured values match $\langle \e^{\Delta I} \rangle$ (solid lines). Above $L=1.5$, the number of measured values corresponding to $k=0$ fall below 100, which is insufficient to estimate $\mean{\e^{-w}}$ with good precision. Error bars correspond to the statistical uncertainty (standard deviation over square root of the sample number).}
\end{figure}

Although multiple measurements of position are involved in this protocol, $\Delta I$ is determined entirely from the index $k$ of the protocol, by a binary criterion: $k=0$ or $k>0$. Using Eqs.~\ref{eq:Delta_I1} and \ref{eq:Delta_Isup1}, we can evaluate $\e^{-w + \Delta I}$ from experimental data, and then take the average of this quantity to test Eq.~\ref{eq:ashida}. Figure~\ref{Fig:JarzFB1}(a) shows these averages for a range of values of $L$ and $h$, revealing agreement with the theoretical prediction. We can also compute $\mean{\e^{-w}}_{k}$ separately for both the $k=0$ and $k>0$ cases and compare it to the prediction of Eq.~\ref{eq:CondAvg_k}, again obtaining agreement as seen in Fig.~\ref{Fig:JarzFB1}(b). Finally, averaging over all the data (rather than separately for $k=0$ and $k>0$), Fig.~\ref{Fig:JarzFB1}(b) reveals excellent experimental verification of Eq.~\ref{eq:newFT}. 

For an engine, the interesting quantity is the mean extracted work, which is upper-bounded by the mean information: $-\mean{w}\leq \mean{\Delta I}$. Fig.~\ref{Fig:bound_W}(a) reveals that, with our first-passage protocol, the mean information provides a tight bound on the mean extracted work. This result is a generic feature of the protocol and of the information definition. The reason is that for large $L+h$ the terms corresponding to $k=0$ in $\mean{\Delta I}$ and $\mean{-w}$ go to zero whereas the terms for $k>1$ go to $2Lh$ in both expressions, saturating in this way the bound. In SM section \ref{SMC} we show that this feature does not depend on the choice of the potential, and results from the sufficient and necessary conditions that each protocol $\Lambda_k$ produces a peaked work distribution. 

We have proposed a framework for obtaining fluctuation theorems and work bounds for information engines powered by error-free measurements. By defining the information-like quantities $I$ and $I_u$ in terms of protocols rather than underlying measurement outcomes, we develop an approach based on time that is more tractable than the fine-grained approach of Ref.~\cite{ashida_general_2014}. We have validated our method experimentally with an engine based on first-passage times, for which the work bound is nearly saturated. The theoretical framework is broader than its application to the current experiment (see EM), and could be applied successfully to first-passage measurements already described in the literature~\cite{Ribezzi_2019,Lee2018}.

\acknowledgments
This work has been partially funded by project ANR-22-CE42-0022. C.J. acknowledges support from the Simons Foundation (award no. 681566) and from the ENS de Lyon. A.I. acknowledges support from the CNRS as an invited researcher in the ENS de Lyon.

The data supporting this study are openly available in Ref.~\onlinecite{Archambault-Zenodo-2025}.

\renewcommand{\theequation}{EM\arabic{equation}}
\renewcommand{\thefigure}{EM\arabic{figure}}
\setcounter{equation}{0}

\onecolumngrid
~
\clearpage

\begin{center}
\large{\bf End Matter}
\end{center}

~
\twocolumngrid

\section{Work distribution and mean value}
In this paper, we use $w$ to denote the work performed on the system, hence $-w$ is the work extracted from it.
Such extraction is performed when the demon is triggered by the condition $x>h$. This can occur when the demon is activated at once, that is when the system is in equilibrium in the potential $U_{-L}(x)=\frac{1}{2}(L+x)^2$ and $x>h$, or it can occur later when the threshold is crossed, thus at $x=h$. The probability distribution function (pdf) of $-w$ is thus
\begin{align}  \label{eq:pdfw}
P(-w)=\ &\frac{1}{2L\sqrt{2\pi}}\exp\left[-\frac{1}{2}\left(\frac{w}{2L}-L\right)^2\right]\theta(-w-2Lh) \nonumber \\
&+\left[1-\frac{1}{2}\erfc\left(\frac{L+h}{\sqrt{2}}\right)\right]\delta(w+2Lh),
\end{align}
where $\theta(-w-w_0)$ is the Heaviside function ($0$ if $-w<w_0$, $1$ otherwise).
An example of this pdf is plotted in Fig.~\ref{Fig:pdf_FB1} for an experiment carried out at $L=0.6$ and $h=0.25$, and it provides an excellent match to the experimental data.

From the pdf, we can compute the mean work:
\begin{equation} \label{eq:meanW}
\begin{split}
\mean{-w}=L \left[2h-(h+L)\erfc\left(\frac{L+h}{\sqrt{2}}\right)\right]\\+\frac{2L}{\sqrt{2\pi}}\exp\left[-\frac{(L+h)^2}{2}\right]
\end{split}
\end{equation}
The resulting prediction is plotted in Fig.~\ref{Fig:bound_W}(a) along the experimental data and the upper bound given by $\mean{\Delta I}$, all in very good agreement.

\section{Rigorous protocol definition}
For the theoretical proofs, we need the protocols to be of the form of Eq.~\ref{eq:defLambda_k}, that is to start with $\lambda_0=A$ and end with $\lambda_M=B$. We therefore add the rule that if all $x_0, x_2\ldots x_{M-1}< h$ then $\lambda$ is switched from $A$ to $B$ at $t = t_M$, regardless of the value of $x_M$. This rule is not implemented experimentally, which is anyway not a problem by choosing a time $t_M$ sufficiently large so that in all practical cases the demon will trig before $t_M$.

\section{Derivation of Eqs.~\ref{eq:Expavg} and \ref{eq:CondAvg_k}} 
By averaging  Eq.~\ref{EqDefDI} we obtain:
\begin{align}
\mean{ \e^{-w+\Delta F-\Delta I}}_\mathrm{FB}
&= \sum_{X, \Lambda_k} \pfb(X, \Lambda_k) \e^{-w+\Delta F} \frac{Q_k}{Q_k^R} \nonumber \\
&= \sum_{\Lambda_k} \frac{Q_k}{Q_k^R} \sum_{X\in\Omega_k} \pno(X\vert\Lambda_k) \e^{-w+\Delta F} \nonumber \\
&= \sum_{\Lambda_k} \frac{Q_k}{Q_k^R} \sum_{X\in\Omega_k} \pRno(X^\dagger\vert\Lambda_k^\dagger)\nonumber \\
&= \sum_{\Lambda_k} Q_k = \sum_{\Lambda_k} P_k = 1, \label{eqproof_ahida}
\end{align}
which is Eq.~\ref{eq:Expavg}. As for Eq.~\ref{eq:CondAvg_k}, we have:
\begin{align}
	\mean{\e^{-w+\Delta F}}_{\mathrm{FB}, k} &= \sum_X \pfb(X\vert \Lambda_k)\e^{-w+\Delta F}
	\nonumber \\
	&= \sum_{X \in \Omega_k} \frac{\pfb(X, \Lambda_k)}{P_k} \e^{-w+\Delta F} \nonumber \\
	&= \frac{1}{P_k} \sum_{X \in \Omega_k} \pno(X\vert\Lambda_k)\e^{-w+\Delta F} \nonumber \\
	\label{eq:EMCondAvg_k}
	&= \frac{1}{Q_k} \sum_{X \in \Omega_k} \pRno(X^\dagger \vert \Lambda_k^\dagger) =  \frac{Q^R_k}{Q_k} = \e^{\Delta I(k)}.
\end{align}

\section{Generalization of Eqs.~\ref{eq:Expavg}-\ref{eq:corrolary} to arbitrary feedback protocols}

We derived Eqs.~\ref{eq:Expavg}-\ref{eq:corrolary} for a protocol involving first-passage times. Here we generalize these results.

Consider a system governed by a Hamiltonian $H(x;\lambda)$, where $x$ denotes the system's microscopic state and $\lambda$ is an externally controlled parameter, which may be continuous or discrete.
The system evolves in time, from $t=0$ to $\tau$, as $\lambda$ is varied according to a schedule, or {\it protocol}, $\Lambda = \{\lambda(t) , t \in [0,\tau] \}$, that is determined by measurement and feedback.
Specifically, measurements are performed at successive times $t_0, \, t_1, \, \cdots \, t_M$, and after the $m$'th measurement, the protocol for varying $\lambda$ from $t=t_m$ to $t_{m+1}$ is determined by all the measurement outcomes up to time $t_m$.
We assume there are no measurement errors, hence the protocol $\Lambda$ is determined uniquely by the system's trajectory $X = \{ x(t) , t \in [0,\tau] \}$.
We assume a discrete set of possible protocols, labelled by the integer $k$.
In $X$-space, let $\Omega_k$ denote the set of all trajectories that produce the protocol $\Lambda_k$.

In the main text, $x \in \mathbb{R}$, $\lambda \in \{A,B \}$, the times $\{t_m\}$ are equally spaced, and the protocols $\Lambda_k$ are determined by first-passage times.
Here we do not make these assumptions.

Under the above-mentioned feedback scheme, the joint probability to obtain $X$ and $\Lambda_k$, and the conditional probability to obtain $X$ when the protocol $\Lambda_k$ is applied without feedback, are related by
\begin{equation}
\label{eq:identity}
P_{\rm FB}(X,\Lambda_k) =
\begin{cases}
P_{\rm no}(X \vert \Lambda_k) \qquad & {\rm if} \quad X \in \Omega_k \\
0 \qquad & {\rm if} \quad X \notin \Omega_k
\end{cases}
\, .
\end{equation}

Now define
\begin{align}
P_k &= \sum_X P_{\rm FB}(X,\Lambda_k) \\
Q_k &= \sum_{X \in \Omega_k} P_{\rm no}(X \vert \Lambda_k) \\
Q_k^R &= \sum_{X \in \Omega_k} P_{\rm no}^R(X^\dagger \vert \Lambda_k^\dagger)
\end{align}
where $^\dagger$ indicates time-reversal, as in the main text.
$P_k$ is the probability to obtain the protocol $\Lambda_k$ when performing the process with feedback;
$Q_k$ is the probability to obtain $X \in \Omega_k$ when protocol $\Lambda_k$ is applied without feedback;
and $Q_k^R$ is the probability to obtain a trajectory $X^\dagger$ whose conjugate twin ($X$) belongs to $\Omega_k$, when protocol $\Lambda_k^\dagger$ is applied without feedback.
By construction, $\sum_k P_k = 1$, and from Eq.~\ref{eq:identity} we have
\begin{equation}
P_k = Q_k \, ,
\end{equation}
hence $\sum_k Q_k = 1$.
However, in general $\sum_k Q_k^R \ne 1$.

Crooks's fluctuation theorem states that
\begin{equation}
P_{\rm no}(X \vert \Lambda_k) \, e^{-w(X,\Lambda_k) + \Delta F} = P_{\rm no}^R(X^\dagger \vert \Lambda_k^\dagger) \, , 
\end{equation}
which combines with Eq.~\ref{eq:identity} and the definition of $Q_k^R$ to give the useful identity,
\begin{equation}
\label{eq:useful}
\sum_{X\in\Omega_k} P_{\rm FB}(X,\Lambda_k) \, e^{-w + \Delta F} = Q_k^R \, .
\end{equation}
Setting $\Delta I(k) = \ln (Q_k^R/Q_k)$ and summing both sides of Eq.~\ref{eq:useful} over $k$ gives
\begin{equation}
\label{eq:corrolary-SM}
\left\langle e^{-w + \Delta F} \right\rangle_{\rm FB} = \sum_k Q_k^R
= \sum_k P_k \, e^{\Delta I(k)}
= \left\langle e^{\Delta I} \right\rangle_{\rm FB}
\end{equation}
If we instead multiply both sides of Eq.~\ref{eq:useful} by $Q_k/Q_k^R$ and then sum over $k$, we get
\begin{equation}
\label{eq:Expavg-SM}
\left\langle e^{-w + \Delta F - \Delta I} \right\rangle_{\rm FB} = \sum_k Q_k = 1
\end{equation}
Finally, dividing both sides of Eq.~\ref{eq:useful} by $P_k$ ($=Q_k$) gives
\begin{equation}
\label{eq:CondAvg_k-SM}
\left\langle e^{-w + \Delta F} \right\rangle_{{\rm FB},k} = \frac{Q_k^R}{Q_k} = e^{\Delta I(k)}
\end{equation}
where $\left\langle \cdots \right\rangle_{{\rm FB},k}$ denotes an average that is conditioned on protocol $\Lambda_k$.
Eqs.~\ref{eq:corrolary-SM}, \ref{eq:Expavg-SM} and \ref{eq:CondAvg_k-SM} correspond, respectively, to Eqs.~\ref{eq:corrolary}, \ref{eq:Expavg} and \ref{eq:CondAvg_k} of the main text.

Following essentially identical steps, we also obtain
\begin{eqnarray}
 \left\langle e^{-w + \Delta F - I} \right\rangle_{\rm FB} &=& \left\langle e^{-I_u} \right\rangle_{\rm FB} \\
 \left\langle e^{-w + \Delta F + I_u} \right\rangle_{\rm FB} &=& \left\langle e^{I} \right\rangle_{\rm FB}.
\end{eqnarray}

In our experiment, we don't have explicit expressions for $I$ and $I_u$ when $k > 0$. It is therefore not straightforward to test those additional fluctuation theorems, but they could prove useful in other cases.

\section{Derivation of $\Delta I(k=0)$,  Eq.~\ref{eq:Delta_I1}} 
Since this measurement occurs in equilibrium, we can compute $Q_0$, $Q_0^R$:
\begin{align}
Q_0 &= P_0 = P(x_0> h) = \int_{h}^{+\infty} \pi_A(x)dx \nonumber \\
&= \int_{h}^{+\infty} \frac{\e^{-\frac{1}{2}(x+L)^2}}{\sqrt{2\pi}} dx = \frac{1}{2}\erfc\left(\frac{h+L}{\sqrt{2}}\right), \label{eq:EMQ0}\\
Q_0^R &= P( X^\dagger \in \Omega_0 \vert \Lambda_0^\dagger) = \int_{h}^{+\infty} \pi_B(x) dx \nonumber \\
&= \int_{h}^{+\infty}\frac{\e^{-\frac{1}{2}(x-L)^2}}{\sqrt{2\pi}} dx = \frac{1}{2}\erfc\left(\frac{h-L}{\sqrt{2}}\right), \label{eq:EMQ0R}
\end{align}
from which we compute $\Delta I(k=0)$ analytically:
\begin{align}
\Delta I(k=0) &= -\ln\left(\frac{Q_0}{Q_0^R}\right) \nonumber \\
&= -\ln\left( \frac{\erfc(\frac{h+L}{\sqrt{2}})}{\erfc(\frac{h-L}{\sqrt{2}})} \right). \label{eq:EMDelta_I1}
\end{align}

\appendix
\setcounter{figure}{0}
\renewcommand{\thefigure}{SM\arabic{figure}}

\onecolumngrid
~
\clearpage

\begin{center}
\large{\bf Supplemental Material}
\end{center}

~

In this Supplemental Material, we provide details on the experimental setup, the derivation of Eq.~\ref{eq:Delta_Isup1}, the computation of the mean information of the SU protocol, and a discussion on the saturation of the information bound on the mean work.

~

\twocolumngrid

\section{Experimental setup} \label{SMA}
\renewcommand{\theequation}{A\arabic{equation}}

\begin{figure}[t]
\centering
\includegraphics[width=\linewidth]{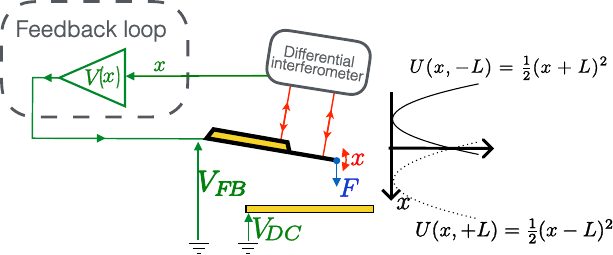}
\caption{ \label{Fig:setup} Experimental setup. The deflection $x$ of a cantilever is measured using an interferometer. $x$ is then used by the feedback loop to compute a voltage $V_\mathit{FB}(x)$ that generates a force on the cantilever, shifting the central position of the harmonic potential.}
\end{figure}

In our experiment a $\SI{1}{mm}$ long conductive cantilever acts as an underdamped mechanical oscillator subject to thermal fluctuations. Fig.~\ref{Fig:setup} sketches our setup, which is similar to the one described in Refs.~\onlinecite{Dago-2021, Dago-2022-JStat, Dago-chapter}. Specifically the first resonant mode of the cantilever is used as a underdamped harmonic oscillator characterized by a stiffness $\kappa\simeq \SI{5e-3}{\newton\per\meter}$, an eigenfrequency $f_0 = \SI{1087}{\hertz}$, a quality factor around $\mathcal{Q}=7$ at atmospheric pressure, and an effective mass $m=\kappa/(2\pi f_0)^2$. The tip deflection $x$ follows the dynamics of a 1D underdamped Brownian particle. The standard deviation of $x$ in thermal equilibrium is $\sigma =\sqrt{k_BT/\kappa}\simeq \SI{0.8}{\nano\meter}$, which is used as the length unit so that all energetic quantities are directly expressed in units of $k_BT$.

The deflection of the cantilever is measured with a quadrature phase interferometer~\cite{paolino_quadrature_2013}. Its outputs are sampled at $\SI{100}{MHz}$ ($\delta t = \SI{10}{\nano\second}$) and processed with a field-programmable gate array device (National Instruments NI FPGA 7975R) that computes the deflection $x$ in real time. The device can be programmed to output a feedback voltage $\vfb$ computed using $x$ and a set of rules implemented by the user. The delay of this feedback is negligible with respect to the oscillator dynamics: it is smaller than $1\si{\micro\second}$, thus three order of magnitude smaller than the period of the oscillator $T_0=1/f_0 \simeq 1\si{\milli \second}$.

The feedback voltage $\vfb$ output by the FPGA is applied to the conductive cantilever, while a constant DC voltage $\vdc = \SI{90}{\volt}$ is applied to a plane electrode about $\SI{500}{\mu m}$ away. This results in a feedback force $F_\mathit{FB}$ on the cantilever:
\begin{equation}
	F_\mathit{FB} \propto (\vdc-\vfb)^2 = ( \vdc^2 - 2\vdc \vfb + \vfb ^2).
\end{equation}
The term $\vdc ^2$ is constant and only shifts the equilibrium position of the oscillator. It is included by defining the origin $x=0$ to the center of the new harmonic potential.
Since the maximum output voltage $\vfb$ possible for the FPGA is $\SI{1}{\volt}$, we can further simplify $F_\mathit{FB}$ noting that $\vfb ^2 \ll \vdc \vfb $. The resulting expression for the force is thus $F_\mathit{FB} \propto 2\vdc \vfb $. The DC bias acts as an amplification factor which is experimentally used to tune the sensitivity of the cantilever to the feedback voltage. 

\section{Derivation of Eq.~\ref{eq:Delta_Isup1}} \label{SMB}
\renewcommand{\theequation}{B\arabic{equation}}

For $k>0$, let $X = (x_0,x_1,\dots,x_M)$ denote a trajectory that belongs to $\Omega_k$. That is, $x_0,x_1,\dots,x_{k-1} < h$ and $x_k>h$. The probability to obtain this trajectory when performing protocol $\Lambda_k$ without feedback is (see Eq.~\ref{pno:def})
\begin{eqnarray}
	\pno(X\vert\Lambda_k)& = &\pi_A(x_0) \cdot p_A(x_1\vert x_0) \cdots \nonumber \\ 
	& & p_A(x_k \vert x_{k-1}) \cdot p_B(x_{k+1} \vert x_k) \nonumber \\
	& & \cdots p_B(x_M\vert x_{M-1}).
 \label{eq:pF}
\end{eqnarray}
The probability to obtain the time-reversed trajectory $X^\dagger$ when performing protocol $\Lambda_k^\dagger$ without feedback is
\begin{eqnarray}
	\pno(X^\dagger\vert\Lambda_k^\dagger)& = &\pi_B(x_M) \cdot p_B(x_{M-1}\vert x_M) \cdots \nonumber \\ 
	& & p_B(x_k \vert x_{k+1}) \cdot p_A(x_{k-1} \vert x_k) \nonumber \\
	& & \cdots p_A(x_0\vert x_1).
 \label{eq:pR}
\end{eqnarray}
Using the detailed balance relation
\begin{equation}
 \frac{p_\lambda(x \vert x^\prime)}{p_\lambda(x^\prime \vert x)} = \frac{\exp[- U_\lambda(x)]}{\exp[-U_\lambda(x^\prime)]}
 \ ,\quad \lambda = A,B,
\end{equation}
and taking the ratio of Eqs.~\ref{eq:pF} and \ref{eq:pR}, we obtain (after the cancellation of many Boltzmann-like factors)
\begin{eqnarray}
 \frac{\pno(X\vert\Lambda_k)}{\pno(X^\dagger\vert\Lambda_k^\dagger)} &=& \exp[ U_B(x_k)-U_A(x_k)] \nonumber \\
 &=& e^{-2Lx_k} \approx e^{-2Lh}.
\end{eqnarray}
On the last line we have assumed that the observation time step $\delta t$ is very short, hence $x_k \approx h$.
Treating this approximation as an equality, we obtain
\begin{eqnarray}
 Q_k &=& \sum_{X\in\Omega_k} \pno(X\vert\Lambda_k) \nonumber \\
 &=& \sum_{X\in\Omega_k} \pno(X^\dagger\vert\Lambda_k^\dagger) e^{-2Lh} = e^{-2Lh} Q_k^R.
\end{eqnarray}
Combining this result with Eq.~\ref{EqDefDI} gives Eq.~\ref{eq:Delta_Isup1}:
\begin{equation}
 \Delta I(k>0) = -\ln \frac{Q_k}{Q_k^R} = 2Lh.
\end{equation}

\section{Mean information of the SU protocol}
\renewcommand{\theequation}{C\arabic{equation}}

\begin{figure}[b]
 \centering
 \includegraphics{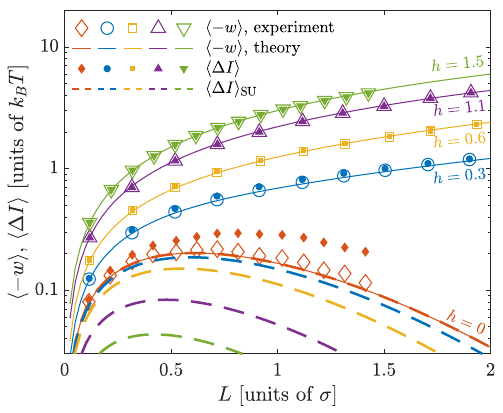}
 \caption{Same as Fig.~\ref{Fig:bound_W}(a), except for the log scale on the vertical axis, and added curves for the mean information of the SU protocol $\mean{\Delta I}_\mathrm{SU}$ (dashed lines). The mean work of our demon, which rapidly samples the system, is always greater than the upper bound of the SU protocol (saturated in this experiment~\cite{Archambault-EPL}).}
 \label{Fig:DI_SU}
\end{figure}

Following Ref.~\onlinecite{ashida_general_2014}, we derive in Ref.~\onlinecite{Archambault-EPL} the following expression of the information $\Delta I_\mathrm{SU}$ valid for the Sagawa-Ueda (SU) protocol:
\begin{equation}
\begin{split}
\text{if } x<h, \quad \Delta I_\mathrm{SU}(x) & = 0\\
\text{if } x>h, \quad \Delta I_\mathrm{SU}(x) & = 2Lx
\end{split}
\end{equation}
where $x$ is the outcome of the measurement in equilibrium in the potential $U(x,-L)$. We easily compute:
\begin{align}
\mean{\Delta I}_\mathrm{SU} &= \int_h^{\infty} 2Lx \frac{1}{\sqrt{2\pi}}\exp\left[-\frac{\left(x+L\right)^2}{2}\right] \dd x \\
& = \frac{2L}{\sqrt{2\pi}} \exp\left[-\frac{\left(h+L\right)^2}{2}\right] -L^2 \erfc\left(\frac{h+L}{\sqrt{2}}\right)
\end{align}
This expression is plotted in Fig.~\ref{Fig:DI_SU} with dashed lines. Unsurprisingly since it describes a different protocol, $\Delta I_\mathrm{SU}$ is unsuited to the rapid sampling demon under scrutiny here.

\section{On the bound of the mean work} \label{SMC}
\renewcommand{\theequation}{D\arabic{equation}}

\begin{figure}[b]
 \centering
 %\vspace{0.5cm}
 \includegraphics{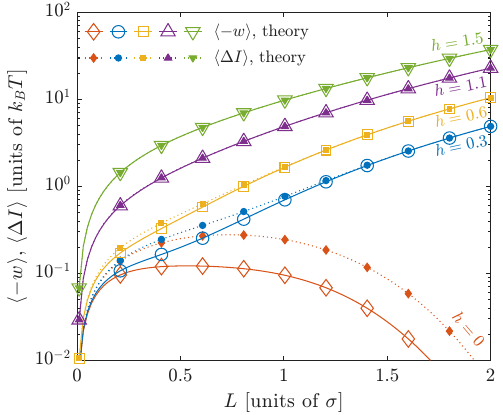}
 \caption{$\mean{-w}$ and $\mean{\Delta I}$ versus $L$ computed from their theoretical expressions for a potential $U_L(x)=\frac{1}{4}(x-L)^4$, for 5 values of $h$ from $h=0$ to $h=1.5$. Results are very similar to Fig.~\ref{Fig:DI_SU} for the harmonic potential, and the bound is saturated for $L\rightarrow 0$ or large $L$ and $h$.}%\vspace{-0.9cm}
 \label{Fig:quartic}
\end{figure}

In Fig.~\ref{Fig:bound_W}(a), $\mean{\Delta I}$ gives a tight bound on $\mean{w}$. This remarkable result is induced by the protocol and by the specific definition of information that we used. If the trigger is not instantaneous ($k>0$), then due to the rapid sampling the extracted work is always the same: $-w (k>0) = U_{B}(h) - U_{A}(h)$. Eq.~\ref{eq:CondAvg_k} then implies that $\Delta I(k>0)= -w(k>0)+\Delta F$, hence the bound is saturated when $k>0$: our information is optimal and we know exactly how much energy we will extract from the system. The deviation from saturation of the bound therefore comes from the contribution $k=0$, which occurs with probability $Q_0$. As long as $Q_0$ is small (few instantaneous triggers), the bound is tight.  In our experiment, large values of $L$ or $h$ induce a vanishing $Q_0$, resulting in the saturation of bound in Fig.~\ref{Fig:bound_W}(a). However this saturation is independent of the shape of the potential (see Fig.~\ref{Fig:quartic} for an example), and a generic property of both the protocol and the definition of information.

Even if $Q_0$ is not small, the bound is saturated in the limit $L \rightarrow 0$. From Eqs.~\ref{eq:EMQ0}-\ref{eq:EMQ0R}, we have
\begin{align}
Q_0 & = \int_{h}^{+\infty} \frac{e^{-U_A(x)}}{Z_A}dx \\
Q_0^R &= \int_{h}^{+\infty} \frac{e^{-U_B(x)}}{Z_B} dx = \int_{h}^{+\infty} \frac{e^{-U_A(x)-\Delta U(x)}}{Z_B} dx,
\end{align}
with $\Delta U = U_B-U_A$. When $L$ is small, so is $\Delta U$, and we linearize $\exp(-\Delta U) \simeq 1-\Delta U$. Since the partition functions $Z_A$ and $Z_B$ are equal, we obtain
\begin{align}
Q_0^R &\simeq \int_{h}^{+\infty} \frac{e^{-U_A(x)}[1-\Delta U(x)]}{Z_A} dx\\
&\simeq Q_0-\int_{h}^{+\infty} \Delta U(x) \frac{e^{-U_A(x)}}{Z_A} dx\\
&\simeq Q_0(1 +\mean{w}_{k=0}).
\end{align}
We therefore have
\begin{equation}
    \Delta I(k=0) = \ln\left(\frac{Q_0^R}{Q_0}\right) \simeq \mean{w}_{k=0}.
\end{equation}
When $L$ is small, the equality between $\Delta I(k)$ and $\mean{w}_{k}$ is valid for any $k$, so the saturation of the bound is reached. Note again that this argument stand for any small transformation of the potential, and doesn't rely on the specific shape we have chosen in our particular experiment.

Thus, for any $h$, the bound is saturated at small and large values of $L$. For the intermediate range of $L$ there are deviations from saturation, but the ratio $R(h,L)=-\langle \Delta I \rangle /\langle w \rangle \rightarrow  1$ for increasing $h$. Specifically $R(h,L) < 1.1$ for all $L$ as long as $h>0.1$ in the case of harmonic potential. These small discrepancies from the saturation cannot be observed on the scale of Fig.~\ref{Fig:bound_W}(a), and are only apparent in  Figs.~\ref{Fig:DI_SU} and \ref{Fig:quartic} due to the log scale on the vertical axis.

From Eq.~\ref{eq:CondAvg_k-SM} established in the general case of a countable protocol-based information $\Delta I(k)$, we can deduce a sufficient and necessary criterion for the information bound to be saturated. Applying Jensen's inequality for each value of $k$ in Eq.~\ref{eq:CondAvg_k-SM}, we see that global saturation results from the saturation of each statistically relevant $k$ (when $P_k$ is not negligible). For a given $k$, such saturation occurs if and only if $w+\Delta I(k)-\Delta F \ll 1$, i.e.\ if the work extracted in each protocol $\Lambda_k$ is peaked around a single value $w=\Delta F -\Delta I(k)$. First-passage protocols avoiding a large contribution of the immediate trigger ($P_0 \ll 1$) fall into this category, but other protocols could be designed to meet this criterion. These general arguments also explain the insensitivity of the saturation to the potential shape, illustrated in Fig.~\ref{Fig:quartic}.

\bibliography{First_passage_information_engine}

\end{document}